\documentclass[aps,twocolumn,superscriptaddress,showpacs,nofootinbib]{revtex4-2}
\usepackage{graphicx}
\usepackage{dcolumn}
\usepackage{bm}
\usepackage{hyperref}
\usepackage{upgreek}
\usepackage{color}
\usepackage{soul}
\usepackage{epstopdf}
\usepackage{units}
\usepackage{longtable}
\usepackage{floatrow}
\usepackage{verbatim} 
\usepackage{latexsym}
\usepackage{gensymb}
\usepackage{floatrow}
\usepackage{amssymb}
\usepackage{ulem}

\makeatletter
\newcommand{\fmarki}{*}
\newcommand{\fmarkii}{*}
\newcommand{\fmarkiii}{\ensuremath{\ddagger}}
\newcommand{\fmarkiv}{\ensuremath{\mathsection}}
\newcommand{\fmarkv}{\ensuremath{\mathparagraph}}
\newcommand{\fmarkvi}{\ensuremath{\|}}
\newcommand{\fmarkvii}{**}
\newcommand{\fmarkviii}{\ensuremath{\dagger\dagger}}
\newcommand{\fmarkix}{\ensuremath{\ddagger\ddagger}}
                
\def\@fnsymbol#1{{\ifcase#1\or \fmarki\or \fmarkii\or \fmarkiii\or \fmarkiv\or \fmarkv\or \fmarkvi\or \fmarkvii\or \fmarkviii\or \fmarkix \else\@ctrerr\fi}}
\makeatother

\begin{document}
\title{Finite field transport response of a dilute magnetic topological insulator based Josephson junction}

\author{Pankaj Mandal}
\let\thefootnote\relax\footnotetext{Corresponding authors:}
\email{pankaj.mandal@physik.uni-wuerzburg.de}
\affiliation{Faculty for Physics and Astronomy (EP3),
Universit\"at W\"urzburg, Am Hubland, D-97074, W\"urzburg, Germany}
\affiliation{Institute for Topological Insulators, Am Hubland, D-97074, W\"urzburg, Germany}

\author{Nicolai Taufertsh\"ofer}
\affiliation{Faculty for Physics and Astronomy (EP3),
Universit\"at W\"urzburg, Am Hubland, D-97074, W\"urzburg, Germany}
\affiliation{Institute for Topological Insulators, Am Hubland, D-97074, W\"urzburg, Germany}

\author{Lukas Lunczer}
\affiliation{Faculty for Physics and Astronomy (EP3),
Universit\"at W\"urzburg, Am Hubland, D-97074, W\"urzburg, Germany}
\affiliation{Institute for Topological Insulators, Am Hubland, D-97074, W\"urzburg, Germany}

\author{Martin P. Stehno}
\affiliation{Faculty for Physics and Astronomy (EP3),
Universit\"at W\"urzburg, Am Hubland, D-97074, W\"urzburg, Germany}
\affiliation{Institute for Topological Insulators, Am Hubland, D-97074, W\"urzburg, Germany}

\author{Charles Gould}
\email{gould@physik.uni-wuerzburg.de}
\affiliation{Faculty for Physics and Astronomy (EP3),
Universit\"at W\"urzburg, Am Hubland, D-97074, W\"urzburg, Germany}
\affiliation{Institute for Topological Insulators, Am Hubland, D-97074, W\"urzburg, Germany}

\author{Laurens W. Molenkamp}
\affiliation{Faculty for Physics and Astronomy (EP3),
Universit\"at W\"urzburg, Am Hubland, D-97074, W\"urzburg, Germany}
\affiliation{Institute for Topological Insulators, Am Hubland, D-97074, W\"urzburg, Germany}

\date{\today}

\begin{abstract}
ABSTRACT: Hybrid samples combining superconductors with magnetic topological insulators are a promising platform for exploring exotic new transport physics. We examine a Josephson junction of such a system, based on the dilute magnetic topological insulator (Hg,Mn)Te and the type II superconductor MoRe. In the zero and very low field limit, to the best of our knowledge, the device shows, for the first time, induced supercurrent through a magnetically doped semiconductor, in this case a topological insulator. At higher fields, a rich and hysteretic magnetoresistance is revealed. Careful analysis shows that the explanation of this behaviour can be found in magnetic flux focusing stemming from the Meissner effect in the superconductor, without invoking any role of proximity induced superconductivity. The phenomena is important, as it will ubiquitously co-exist with any exotic new physics that may be present in this class of devices.


\end{abstract}

\maketitle

The combination of superconductivity with chiral one dimensional edge modes is expected to reveal intriguing transport properties\cite{ma1993josephson, van2011spin, gavensky2021nonequilibrium} as well as to potentially host chiral Majorana modes\cite{qi2010chiral, gavensky2020majorana, chaudhary2020vortex}. A possible way forward is to exploit quantum anomalous Hall systems and their quantum edge channels at zero magnetic field\cite{yu2010quantized, chang2013experimental, checkelsky2014trajectory, chang2015high, bestwick2015precise, grauer2015coincidence}. However, the strong internal fields in these materials may hamper the observation of a supercurrent. Indeed attempts in this direction thus far have reported Andreev reflection at the superconductor to ferromagnetic topological insulator interface, but no evidence of a supercurrent\cite{kayyalha2020absence}.

Alternatively, appropriate hybrid devices have therefore recently attracted much research interest, with experimental efforts in various systems and device geometries\cite{wan2015induced, amet2016supercurrent, lee2017inducing, seredinski2019quantum, wei2019chiral}. A significant challenge however, is that the formation of chiral edge modes generally requires a large magnetic field (of the order of a few Tesla), which tends to suppress any proximity induced superconductivity.

For this reason, a promising path to merge superconductors with chiral edge states is to make use of dilute magnetic topological insulators. Manganese doped HgTe quantum wells are two dimensional topological insulators where, at lowest temperatures, the spin Hall conductance is quantized due to Kondo screening of the magnetic impurities\cite{shamim2021quantized}, and they have been demonstrated to show chiral quantum Hall edge modes at fields as low as 50 mT\cite{shamim2020emergent}. Pairing such systems with superconductors should create an ideal platform in which to search for exotic emergent phenomena of transport properties. 

\begin{figure}[h]%
\includegraphics[width=\columnwidth]{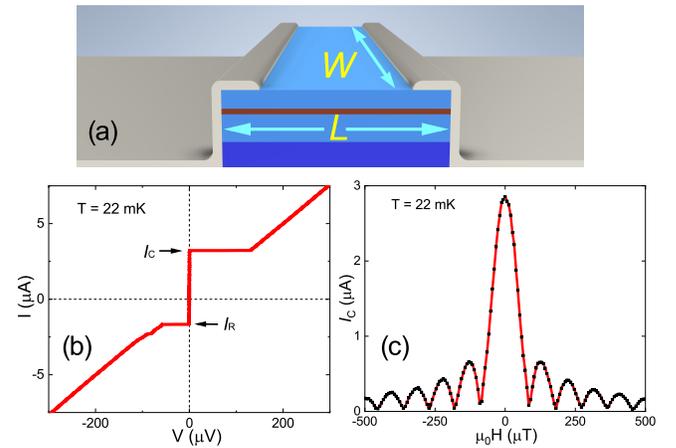}
\caption{(a) Schematic of the side contact geometry. The Mn doped quantum well (red), is sandwiched between (Hg,Cd)Te barrier layers (light blue) grown on a (Cd,Zn)Te substrate (dark blue). (b) I(V) characteristics of a Josephson junction at zero magnetic field, indicating critical current ($I_{C}$) and retraping current ($I_{R}$). (c) Fraunhofer-like diffraction pattern of $I_{C}$($\mu_{0}$H). Measurements are done at 22 mK.  }%
\label{Fig1}%
\end{figure}

In this paper, we report on a transport study of Josephson junction (JJ) devices fabricated from this material system, and which, for the first time, shows that supercurrent can be induced in a dilute magnetic semiconductor (DMS). DMS materials are semiconductors, most typically II-VI materials, containing a few percent of magnetic atoms and have been the subject of intense study for over half a century. The transport physics is well characterized, and while they are known to exhibit a large Zeeman effect due to exchange interaction between the magnetic ions and the semiconductor band structure, they are paramagnets with no remanent magnetization\cite{furdyna1988diluted, kossut1993diluted, dietl2014dilute}.

The weak links in our devices are comprised of an 11 nm thick (Hg,Mn)Te quantum well with a Mn concentration of 2.3\% grown by molecular beam epitaxy. These parameters put it in the topological regime\cite{shamim2020emergent}. It has a carrier density of 8.8x10$^{11}$ cm$^{-2}$ and a mobility of 111 000 cm$^{2}$/Vs, as extracted from Hall measurement on a 180x540 $\mu$m Hall-bar patterned from the same wafer. This material is fashioned into JJ devices by first patterning mesa structures using wet-etching techniques\cite{bendias2018high}. The width $W$ of these mesa is 5.5 $\mu$m, and the length $L$ is varied between 550 and 1200 nm. The mesa is then side-contacted by 80 nm thick MoRe leads, as shown in Fig. 1(a). MoRe was selected for the superconducting electrodes due to its relatively high $T_{c}$ ($\sim$ 9.6 K) and large critical field $H_{c2}$ ($\sim$ 8 T), which allows measurement on a sample with superconducting contacts at relatively high fields\cite{amet2016supercurrent, wei2019chiral, seredinski2019quantum}. The stripe shaped electrodes are  4 $\mu$m wide, and extend for at least 15 $\mu$m before widening to form contact pads. A schematic of the resulting device geometry, including a labelling of the device dimensions, is shown in the inset of Fig. 2(a).

As seen in Fig. 1(b), for the 850 nm long device, when cooled to the base temperature of a dilution fridge, and without an applied magnetic field, the device shows a strong induced supercurrent of about 3.2 $\mu$A. This critical current varies under a small perpendicular-to-plane magnetic field following the expected Fraunhofer pattern shown in Fig. 1(c). The x-scale in Fig. 1(c) has been shifted by about 13 $\mu$T to correct for the remanent field of the magnet. Note that the field range used here is below $H_{c1}$ of the MoRe which is about 1 mT.

As is well known\cite{tinkham2004introduction}, the period of the Fraunhofer pattern  gives the effective area from which flux lines are threaded into the junction. This area $A_{C}$ collects field lines not only from the nominal area of the junction (given by $L \cdot W$), but, because of the Meissner effect, also some field lines which impinge on the superconducting leads and are then focused through the junction. The constant magnetic field spacing of the minima in the Fraunhofer pattern of Fig. 1(c) indicate that the superconducting leads are in a perfect Meissner state, and thus able to fully expel the magnetic field\cite{suominen2017anomalous}. In this regime, the collection area can be determined from the device geometry. Roughly speaking, field lines take the shortest route possible around the superconductor. For the case of stripe shaped contacts running across the edge of the mesa, as sketched in Fig. 2(a), the junction then collects from an approximately rectangular area having the same width as the junction, and extending about half of the 4 $\mu$m width into each superconducting stripe\cite{amet2016supercurrent}, giving an area of 22 $\mu$m$^2$. This collection area is indicated by the dashed outline in the schematic. Note that the definition of $L^*$ in Fig 2(a) is 300 nm less than $L$ due to overlap between the superconducting leads and the mesa. Removing this 1.65 $\mu$m$^2$ overlap region from the 22 $\mu$m$^2$ gives an area approximately 20 $\mu$m$^2$.

The period of 90 $\mu$T observed in Fig. 1(c) corresponds to $A_{C}$  = 23 $\mu$m$^2$, in reasonable agreement with the area indicated in Fig. 2(a). Other junctions with lengths of 550 and 1200 nm show qualitatively similar behaviour, and have periods of 96 and 86 $\mu$T, corresponding to $A_{C}$ = 21.6 and 24.0 $\mu$m$^2$, respectively. These values are consistent with the geometric argument given above. They show that the effective area of each of the junction is increased by about 18 $\mu$m$^2$ due to the focusing, which compares well to the 20 $\mu$m$^2$.

\begin{figure}[h]%
\includegraphics[width=\columnwidth]{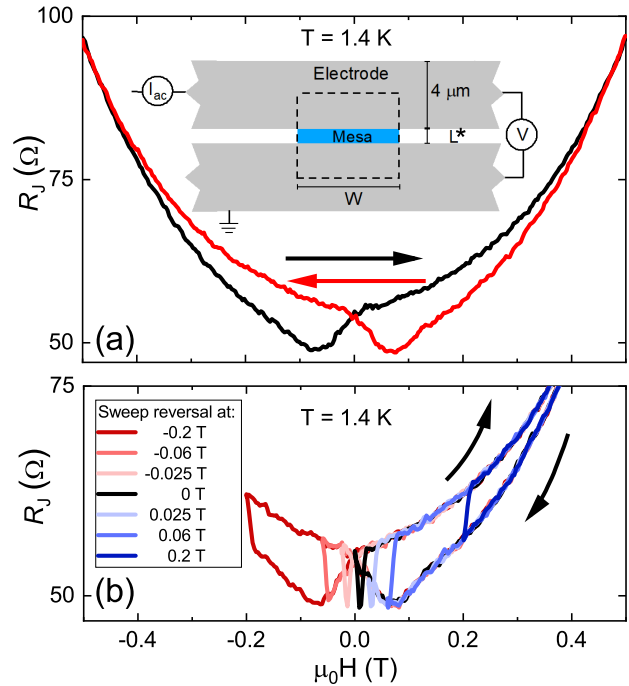}
\caption{ (a) Resistance of the Josephson junction as a function of $\mu_{0}H$. Inset: Schematic of the junction geometry and quasi-four probe measurement configuration. The dashed rectangle indicated the field collection area $A_{C}$. (b) Minor loops of the magnetoresistance: Starting from large positive values, the field is first swept towards negative values. The sweep direction is then reversed at -0.2 T, -0.06 T, -0.025 T, 0 T, 0.025 T, 0.06 T and 0.2 T, and in each case the resistance value immediately jumps to the other branch of the anti-hysteresis. All measurements are done at 1.4 K, and the arrows indicate the sweep directions.}%
\label{Fig2}%
\end{figure}

These base temperature, low field measurements confirm that our device constitutes a proper Josephson junction, with robust supercurrent flowing through a magnetically doped topological insulator. Apparently, dilute concentration of the magnetic impurities and the Kondo screening reported in\cite{shamim2021quantized}, allows for induced superconductivity. In order to achieve the previously discussed goal of pairing superconductivity with chiral edge states, a larger field of at least order 50 mT will be needed\cite{shamim2020emergent}. When such a larger magnetic field is applied however, the resistance of the sample shows some unexpected characteristics, which do result from the magnetic doping in the weak link. Specifically, as the field range of the scan is increased, a strong magnetoresistance is observed but with a minimum of resistance which is not at zero field, and has a position that depends on field history. Note that, as seen in Fig. 1(c), the oscillation of the supercurrent has already decayed by more than a factor of 15 at 0.5 mT compared to the zero-field value. As such it can be fully neglected on the scale of magnetoresistance feature which show a shift in resistance minimum at approximately 80 mT.

We systematically explore these characteristics in a study of the magneto-transport properties of these JJ, conducted at a temperature of 1.4 K, where the MoRe leads are superconducting, but any supercurrent in the topological insulator is excluded. We determine the device resistance $R_{J}$ in the presence of a perpendicular-to-plane magnetic field $\mu_{0}$H using the four-probe measurement configuration indicated in the inset of Fig. 2(a), and standard low frequency lock-in techniques. Such a measurement is shown in Fig. 2(a) for a junction with $L$ = 850 nm. It shows what one might term an ‘‘anti-hysteretic’’ behaviour as a function of applied field. That is to say that while the overall behaviour of the device shows positive magnetoresistance, the minimum in resistance occurs before the field reaches zero in either sweep direction.

This unusual behaviour becomes even more curious in minor loops as shown in Fig. 2(b). Here we start from +0.5 T and sweep the field down to various points, stop, and reverse the sweep direction back towards +0.5 T. In each case, the change in sweep direction leads to an abrupt jump of the device resistance to the opposite branch of the anti-hysteretic curve. A qualitatively similar behaviour is also observed in JJ with lengths of 550 nm and 1200 nm (not shown). For the three curves for $\pm$0.025 T and 0 T in Fig. 2(b) for which the turnaround behave differently to the other curves in that they show a down and up jump instead of a single jump to the other branch. This somewhat distinct behaviour is discussed below.

While it may be tempting to assign this unusual magnetoresistive and sweep direction dependent behaviour to the topological nature of the material system under investigation, one must first determine whether this rich magnetotransport phenomenology can stem from more classical influences of the superconducting leads, such as the magnetic flux focusing attributed to the Meissner effect.

To examine this possibility, we fabricate devices of the same geometry as above, but with non-superconducting Ti/Au contacts instead of the MoRe, and repeat the same measurements as in Fig. 2. The result is shown in Fig. 3(a) for a 850 nm long junction with normal contacts. This device exhibits a roughly parabolic, positive magnetoresistance of the type usually explained by the effect of the Lorentz force on the electrons, and with no sign of hysteresis. The measured resistance includes the sheet resistance of the (Hg,Mn)Te plus the contact resistance of the Ti/Au leads, which can be assumed to be independent of magnetic field. From the known zero-field sheet resistance ($\sim$ 70 $\Omega$/$\square$) of this quantum well material, and the device geometry, the contact resistance can be estimated, and subtracted from the curve, leaving only the actual magnetoresistance ($MR$) of the material. This is shown as the blue curve in Fig. 3(b), with the data normalized to the zero field value. This result can be viewed as a calibration curve for using the (Hg,Mn)Te layer as a magnetic field sensor in the sense that the amount of magnetoresistance in the material directly indicates the amplitude of the magnetic field to which it is exposed.

\begin{figure}[h]%
\includegraphics[width=\columnwidth]{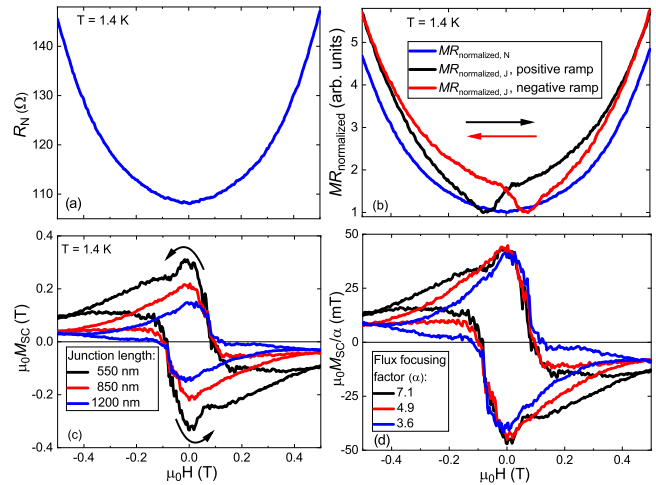}
\caption{ (a) Resistance of a 850 nm long normal metal junction as a function of applied magnetic field. (b) Normalized magnetoresistance for the normal and superconducting junction device after the contract resistance was removed (see text for details). (c) and (d) Plots of the effective magnetization contribution to the field sensed in the superconducting junction, extracted from the transport measurement as described in the text. (c) gives the raw extracted values, while in (d) they are normalized by the flux focusing amplification ratio.}%
\label{Fig3}%
\end{figure}

To compare the results from our JJ devices with this curve, we follow the same process for removing the contact resistance in the case of the superconducting devices. Here we assume that the field at which the resistance is minimal is the effective zero field condition, and normalize the curves to that value. The results are the black and red curves in Fig. 3(b). 

The total magnetic field to which the (Hg,Mn)Te is subjected can be described by a superposition of the applied external field plus an effective magnetization $M_{SC}$ contribution stemming from the flux focusing effect of the superconducting leads: $B_{eff}$ = $\mu_{0}M_{SC}$ + $\mu_{0}H$. $M_{SC}$ as a function of applied field can then be extracted by comparing the curves from Fig. 3(b) for the two device types. The result of this analysis for JJ of three different lengths is given in Fig. 3(c).

The three curves are similar in overall shape and differ only in the strength of the effective magnetization contribution. Keeping in mind that the Meissner effect focuses field from the collection area $A_{C}$ into the junction area, we normalize the curves by a flux amplification ratio $\alpha$ = $A_{C}$/$L \cdot W$ determined from the Fraunhofer pattern, and show the results in Fig. 3(d).

The curves roughly collapse onto each other, confirming that flux focusing plays a key role in describing the magnetoresistance of the junctions. The deviation at larger field for the 550 nm device is likely due to uncertainty in the exact device dimensions. Indeed, the value of the black curve at 0.5 T can be brought into agreement with the other two curves by adjusting the nominal length of the shortest junction, and thus the corresponding materials resistance, by about 20\%, which is within the lithographic accuracy of the wet etching process used to define the mesa. 

To further confirm that the magnetoresistance behaviour results directly from the field expulsion properties of the superconductor, we study the magnetic response of the superconducting film in a SQUID. We take a piece of the same wafer as used for the devices and etch away the quantum well to provide the same surface as in the actual devices. We then evaporate a MoRe film using the same process parameters as for the superconducting lead, and mount a 3.0 x 3.1 mm$^2$ piece of this layer into a SQUID magnetometer.

We measure the magnetic moment ($m$) of this layer in response to a magnetic field applied perpendicular to the layer plane. A reference measurement is done at 15 K, above the $T_{c}$ of MoRe, to determine the diamagnetic contribution of the substrate and the sample holder. This diamagnetic background is subtracted from the subsequent measurements, which are shown here. The sample is cooled below its $T_{c}$ in zero magnetic field.

Figure 4(a) shows the sample magnetization at 2 K as the magnetic field is swept from -0.5 T to 0.5 T and back. The observed hysteretic behaviour is typical of a hard type II superconductor with strong pinning potential\cite{bean1964magnetization, kim1963magnetization}. Equally interesting are the minor loops presented in Fig. 4(b), where, as in the case of the transport data, a change in sweep direction causes an immediate jump to the opposite branch of the hysteresis curve.

\begin{figure}[h]%
\includegraphics[width=\columnwidth]{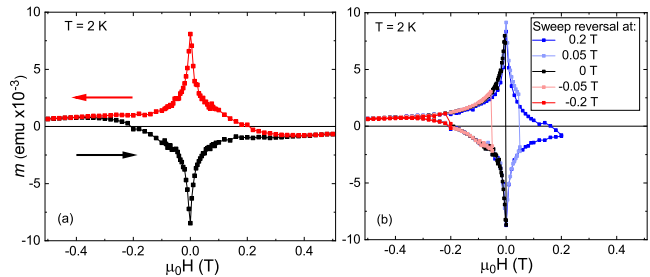}
\caption{(a) SQUID measurement of the magnetization \textit{m} of MoRe in response to a perpendicular-to-plane magnetic field. The measurements are done at 2 K. The field is swept from -0.5 T to +0.5 T and back, with arrows indicating the direction. (b) Minor loops: In five separate curves, the field is swept from -0.5 T to: -0.2 T, -0.05 T, 0 T, 0.05 T and 0.2 T, and then back to -0.5 T.}%
\label{Fig4}%
\end{figure}

This hysteretic behaviour can be understood with the help of Bean's critical-state model\cite{bean1964magnetization}. The basic concept can be described in a toy model: Imagine a hollow superconducting cylinder with the magnetic field applied along its axis. As the field is initially increased, a Meissner current develops in the cylinder and opposes this field. Once the applied field reaches the lower critical field $H_{c1}$ it can no longer be fully screened and some flux lines penetrate through the cylinder walls and into the hollow interior. As the applied field is then decreased, the sign of the screening current on the outer surface changes to again oppose a change in field. Some field lines become trapped in the interior, and remain once the field is returned to zero, contributing a remanent magnetization. In real samples, the role of the hollow part of the cylinder is played by strong pinning centres. Since the SQUID detects contributions from both the trapped flux, as well as the field induced by the screening currents, the Bean model explains well both the hysteresis of Fig. 4(a) and the abrupt jump between curves when the sign of the screening current inverts in response to the changes in sweep direction\cite{shantsev2000thin}.

In the original measurements by Bean on a macroscopic rod, the hysteresis loop was much flatter than the one observed here, and did not show a peak near zero field. Such a low field feature is however known from measurements on thin films of high-$T_{c}$ superconductors \cite{shantsev1999central}. For magnetic fields perpendicular to a thin layer, the large geometric demagnetization factor leads to very small effective $H_{c1}$\cite{fetter1967mixed}. This causes the critical current density to vary strongly with applied field \cite{kim1963magnetization} and results in a sharp maximum in effective magnetization near zero field.

The overall shape of this magnetization behaviour shows a strong resemblance to the effective magnetization extracted from the transport data. The somewhat broader shape of the 0 T peak in transport compared to SQUID is probably due to the side contact geometry in the device giving a small section of superconductor with a different demagnetization factor, as well as other details of the contacts shape and edge effects. 

The Bean model also allows us to understand the distinct behaviour of the $\pm$0.025 T and 0 T curves from Fig. 2(b). On a magnetic field ramp from high field, the magnetoresistance minimum corresponds to the applied field for which $B_{eff}$ is zero. Once the magnetic field ramp crosses over this zero-$B_{eff}$ region, the device is required to cross through zero-$B_{eff}$ again, once the ramp direction is reversed. This down and up jump is so sharp that it has the shape of a spike, due to the strong susceptibility of the superconductor to flux pinning at low magnetic field.

In summary, we have presented a rich magnetotransport phenomenology observed in Josephson junctions made from a dilute magnetic topological insulator between two MoRe superconducting leads. At lowest temperatures, the junctions show a clear induced supercurrent with a Fraunhofer pattern in very low magnetic fields. At higher applied fields, the magnetoresistance response appears to show exotic behaviour. We demonstrate that this response can be described using established properties of the superconducting leads. Indeed, the transport properties are driven by magnetic flux focusing resulting from the Meissner-like properties and screening current effects within the hard superconductor. Such phenomenology will exist in any junction where type II superconducting leads are coupled to a material showing significant magnetoresistance, with the topology of the material system having no relevance to the results presented here.

This paper aims to raise awareness of this phenomenology, as it is likely to occur in some form in all devices attempting to combine superconductivity with chiral edge modes. A proper understanding of these magnetic flux focusing effects is therefore a key prerequisite to being able to identify any truly exotic transport physics that research into such devices may reveal.

\begin{acknowledgments}
The authors thank T.M. Klapwijk and M. Sawicki for useful discussions. We gratefully acknowledge financial support of the ENB Graduate school on ‘Topological Insulators’, the EU ERC-AG Program (project 4-TOPS), the Free State of Bavaria (the Institute for Topological Insulators), Deutsche Forschungsgemeinschaft (SFB 1170, 258499086), and the W\"urzburg-Dresden Cluster of Excellence on Complexity and Topology in Quantum Matter (EXC 2147, 39085490). 
\end{acknowledgments}

\end{document}